\documentclass{article}
\usepackage{graphicx}

\title{Correlational properties of two-dimensional solvable chaos on the
unit circle}
\author{Aki-Hiro Sato and Ken Umeno \\
Department of Applied Mathematics and Physics, \\
Graduate School of Informatics, Kyoto University, \\
Yoshida-Honcho, Sakyo-ku, 606-8501, Kyoto JAPAN
}
\date{\empty}
\begin{document}
\maketitle
\begin{abstract}
This article investigates correlational properties of two-dimensional
chaotic maps on the unit circle. We give analytical forms of
higher-order covariances. We derive the characteristic function of their
simultaneous and lagged ergodic densities. We found that these
characteristic functions are described by three types of two-dimensional 
Bessel functions. Higher-order covariances between x and y and those
between y and y show non-positive values. Asymmetric features between
cosine and sine functions are elucidated.
\end{abstract}
\section{Introduction}
Knowledge on solvable chaos is useful for designing random number
generators~\cite{Ulam,Adler,Gonzalez:06,Umeno} and Monte Carlo
integration~\cite{Umeno:00}. The idea of applying chaos theory to
randomness has produced important works
recently~\cite{Umeno:98,Umeno:97,Chen,Umeno:06}. Geisel 
and Fairen analyzed statistical properties of Chebyshev
maps~\cite{Geisel}. They showed the mixing 
properties and higher order moments with higher-order characteristic
functions. Gonz\'alez and Pino proposed a pseudo random number generator
based on logistic maps~\cite{Gonzalez:99}. Collins et al.~\cite{Collins}
have applied the logit transformation to the logistic map variable for
producing a sequence with a near Gaussian distribution. These solvable
chaotic properties enable us to design and employ chaos for application
purposes.

First, let us consider maps in the form of Chebyshev polynomials of degree
$k$
\begin{equation}
x_{t+1} = T_k(t_t),
\label{eq:Chebyshev}
\end{equation}
which map the interval $[-1,1]$ onto the same interval. The first few
polynomials are explicitly $T_1(x)=x$, $T_2(x)=2x^2-1$, and $T_3(x)=4x^3
- 3x$. Since, there is permutability of the Chebyshev polynomials,
$T_k(T_l(x))=T_{kl}(x)$, Eq. (\ref{eq:Chebyshev}) can be expressed as
\begin{equation}
x_t = T_{k^t}(x_0).
\end{equation}
It was shown by Adler and Rivlin that Chebyshev maps with $k \geq 2$ are
ergodic and strongly mixing. This map dynamics has the invariant measure
$\mu(\mbox{d}x)=\frac{\mbox{d}x}{\pi\sqrt{1-x^2}}$. Geisel and Fairen
shows that the characteristic function of the Chebyshev maps can be expressed as Bessel
function~\cite{Geisel}. They further considered the higher-order characteristic
function. Following their strategy, we consider the characteristic
function of two-dimensional solvable chaotic maps on a unit circle.
We further calculate the higher-order covariance based on the
characteristic function.

This article is organized as follows. In Sec. \ref{sec:2d}, we introduce 
two-dimensional chaotic maps on a unit circle. In
Sec. \ref{sec:sim-cov}, we show that simultaneous covariance among two
variables is independent. In Sec. \ref{sec:sim-high}, we derive
an analytical form of higher-order covariance among two variables. In
Sec. \ref{sec:lag-high}, we compute higher-order covariance among two
variables with lags. Sec. \ref{sec:conclusion} is devoted to concluding
remarks.

\section{Two-dimensional solvable chaos}
\label{sec:2d}

In this article, we consider two-dimensional maps on a unit
circle. Suppose that $z_t=x_t + \sqrt{-1}y_t$ denotes a complex number, where
$x_t$ is a real number and $y_t$ is an imaginary part at step $t \quad
(t=0,1,\ldots)$. Then, we define the complex dynamics as
\begin{equation}
z_{t+1} = z_t^k,
\label{eq:2dmap}
\end{equation}
where $k$ is an integer. We can also express Eq. (\ref{eq:2dmap}) as
\begin{equation}
\left\{
\begin{array}{lcl}
x_{t+1} &=& P_k(x_t,y_t) \\
y_{t+1} &=& Q_k(x_t,y_t)
\end{array}
\right.,
\label{eq:map}
\end{equation}
where $P_k(x,y)$ and $Q_k(x,y)$ are defined as
\begin{eqnarray}
(x+\sqrt{-1}y)^k &=& P_k(x,y) + \sqrt{-1}Q_k(x,y), 
\label{eq:f-k}
\\
x^2 + y^2 &=& 1.
\end{eqnarray}
The first few polynomials are explicitly given by $P_1(x,y)=x$,
$Q_1(x,y)=y$, $P_2(x,y)=x^2-y^2$, $Q_2(x,y)=2xy$, $P_3(x,y)=x^3
- 3xy^2$, $Q_3(x,y)=3x^2y - y^3$, $P_4(x,y)=x^4 - 6x^2y^2 +
y^4$, $Q_4(x,y)=4x^3y - 4xy^3$, $P_5(x,y)=x^5 - 10x^3y^2 + 5xy^4$,
$Q_5(x,y) = 5x^4y-10x^2y^3 + y^5$, $P_6(x,y)=x^6 -
15x^4y^2+15x^2y^4-y^6$, $Q_6(x,y)=6x^5y - 20x^3y^3 + 6xy^5$,
$P_7(x,y)=x^7 - 21x^5y^2 + 35x^3y^4 - 7xy^6$, and 
$Q_7(x,y)=7x^6y - 35x^4y^3 + 21x^2y^5 - x^7$.

In general, $P_k(x,\pm\sqrt{1-x^2}) = T_k(x)$ is satisfied. Specifically,
$Q_k(x,y)$ for odd ordered $k$ is equivalent to
$Q_k(\pm\sqrt{1-y^2},y)=-T_k(y)$.

\begin{figure}[!ht]
\centering
\includegraphics[scale=0.8]{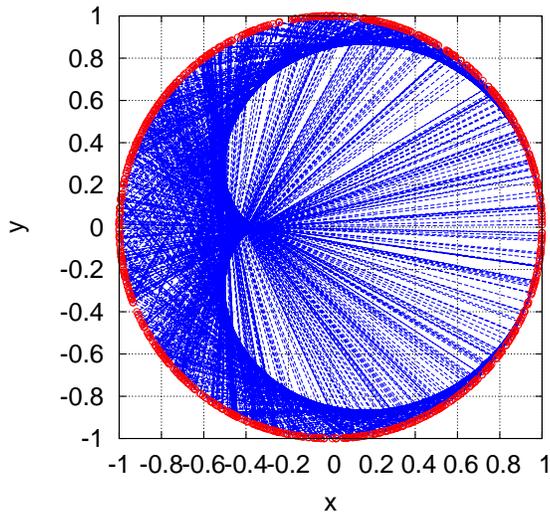}
\caption{800 steps of a trajectory of the two-dimensional chaotic map for
 $k=2$. The initial value is given by $(x_0, y_0)=(-0.820000, 0.572364)$.}
\label{fig:2dmap}
\end{figure}

If we set an initial condition $z_0 = x_0 + \sqrt{-1}y_0$ on the unit
circle $|z_0|=1$, $z_t$ is also mapped on the unit circle. In this case,
Eq. (\ref{eq:f-k}) can be rewritten as
\begin{equation}
\exp(\sqrt{-1}\theta)^k = \exp(k\theta\sqrt{-1}),
\end{equation}
where $\theta$ denotes the argument of $(x,y)$ on the two-dimensional
plane. It is convenient to represent the polynomial $P_k(x,y)$ and
$Q_k(x,y)$ in the form
\begin{equation}
\left\{
\begin{array}{lcl}
P_k(\cos\theta, \sin\theta) &=& \cos(k\theta) \\
Q_k(\cos\theta, \sin\theta) &=& \sin(k\theta)
\end{array}
\right.
\end{equation}
Fig. \ref{fig:2dmap} shows a trajectory of $(x_t,y_t)$ for $k=2$. The
value at each step stands on the unit circle.

By introducing $\theta_t$ as the argument of $z_t$, we have
\begin{equation}
\theta_{t+1} = k\theta_t.
\label{eq:map-theta}
\end{equation}
The solution of Eq. (\ref{eq:map-theta}) can be written as
\begin{equation}
\theta_t = k^t\theta_0,
\label{eq:solution-theta}
\end{equation}
by using $\theta_0$, denoted as the argument of $z_0$. Therefore, $z_t =
x_t + \sqrt{-1}y_t$ is rewritten as
\begin{equation}
z_t = \cos(k^t\theta_0) + \sqrt{-1}\sin(k^t\theta_0) =
 \exp(k^t\theta_0\sqrt{-1}).
\end{equation}

Eq. (\ref{eq:solution-theta}) is ergodic and has the constant invariant density
$\rho_{\Theta}(\theta)=\frac{1}{2\pi} \quad (0 \leq \theta \leq 2\pi)$
since Eq. (\ref{eq:map-theta}) is a Bernoulli map on mod $2\pi$.

Transforming the orthogonal coordinate $(x,y)$ into the polar coordinate
$(r,\theta)$ by $x = r\cos\theta$ and $y = r\sin\theta$, we have
$\rho_R(r)= \delta(r-1)$. Therefore, the joint invariant density of $x$
and $y$ can be described as
\begin{equation}
\rho_{XY}(x,y) =
 \rho_{\Theta}(\theta)\rho_R(r)\Bigl|\frac{\partial(\theta,r)}{\partial(x,y)}\Bigr|
 = \frac{\delta(\sqrt{x^2+y^2}-1)}{2\pi\sqrt{x^2+y^2}},
\label{eq:rhoxy}
\end{equation}
where $\delta(\cdot)$ represents Dirac's $\delta$-function. The marginal
density in terms of $x$ is given by
\begin{eqnarray}
\nonumber
\rho_X(x) &=& \int_{-1}^{1}\rho_{XY}(x,y)\mbox{d}y \\
\nonumber
          &=&
	   \frac{1}{2\pi}\int_{-1}^{1}\frac{\delta(\sqrt{x^2+y^2}-1)}{\sqrt{x^2+y^2}} \mbox{d}y\\
\nonumber
&=&
 \frac{1}{\pi}\int_{|x|-1}^{\sqrt{x^2+1}-1}\frac{\delta(t)}{\sqrt{(t+1)^2-x^2}}\mbox{d}t \\
\nonumber
&=& \frac{1}{\pi\sqrt{1-x^2}}.
\end{eqnarray}
In the same way, we obtain
\begin{equation}
\rho_Y(y) = \int_{-1}^{1}\rho_{XY}(x,y)\mbox{d}x = \frac{1}{\pi\sqrt{1-y^2}}.
\end{equation}
Note that $\rho_X(x)$ and $\rho_Y(y)$ are the same as the ergodic
density of the Chebyshev maps.

\section{Simultaneous covariance}
\label{sec:sim-cov}

Next, let us consider auto-correlations of $x$ and $y$ and
cross-correlation between $x$ and $y$. Obviously, mean values of $x$ and
$y$ are given as zero.
\begin{eqnarray}
\lim_{T\rightarrow\infty}\frac{1}{T}\sum_{t=0}^{T-1}x_t &=&
 \int_{-1}^{1}x \rho_X(x)\mbox{d}x = \int_{-1}^{1}\frac{x}{\pi\sqrt{1-x^2}}\mbox{d}x =
 0, \\
\lim_{T\rightarrow\infty}\frac{1}{T}\sum_{t=0}^{T-1}y_t &=&
 \int_{-1}^{1}y \rho_Y(y)\mbox{d}y = \int_{-1}^{1}\frac{y}{\pi\sqrt{1-y^2}}\mbox{d}y =
 0. 
\end{eqnarray}
We shall introduce four types of correlations:
\begin{eqnarray}
\nonumber
c_{XX}(\tau) &=& \lim_{T\rightarrow
 \infty}\frac{1}{T}\sum_{t=0}^{T-1}x_t x_{t+\tau} = \int_{-1}^{1}\mbox{d}x
 \int_{-1}^{1} \mbox{d}y x \underbrace{P_k\circ\cdots\circ
 P_k}_{\tau}(x,y)\rho_{XY}(x,y) \\
\label{eq:cxx}
\\
\nonumber
c_{YY}(\tau) &=& \lim_{T\rightarrow
 \infty}\frac{1}{T}\sum_{t=0}^{T-1}y_t y_{t+\tau} = \int_{-1}^{1}\mbox{d}x
 \int_{-1}^{1} \mbox{d}y y \underbrace{Q_k\circ\cdots\circ
 Q_k}_{\tau}(x,y)\rho_{XY}(x,y) \\
\label{eq:cyy}
\\
\nonumber
c_{XY}(\tau) &=& \lim_{T\rightarrow
 \infty}\frac{1}{T}\sum_{t=0}^{T-1}x_t y_{t+\tau} = \int_{-1}^{1}\mbox{d}x
 \int_{-1}^{1} \mbox{d}y x \underbrace{Q_k\circ\cdots\circ
 Q_k}_{\tau}(x,y)\rho_{XY}(x,y) \\
\label{eq:cxy}
\\
\nonumber
c_{YX}(\tau) &=& \lim_{T\rightarrow
 \infty}\frac{1}{T}\sum_{t=0}^{T-1}y_t x_{t+\tau} = \int_{-1}^{1}\mbox{d}x
 \int_{-1}^{1} \mbox{d}y y \underbrace{P_k\circ\cdots\circ
 P_k}_{\tau}(x,y)\rho_{XY}(x,y) \\
\label{eq:cyx}
\end{eqnarray}

Transforming the orthogonal coordinate $(x,y)$ into the polar coordinate
$(r,\theta)$, we can calculate Eqs. (\ref{eq:cxx}) to (\ref{eq:cyx}) as
\begin{eqnarray}
c_{XX}(\tau) &=& \frac{1}{2\pi}\int_{0}^{2\pi}\cos\theta \cos
 k^{\tau}\theta \mbox{d}\theta = \frac{1}{2}\delta_{1,k^{\tau}} 
\label{eq:cxx2}
\\
c_{YY}(\tau) &=& \frac{1}{2\pi}\int_{0}^{2\pi}\sin\theta \sin
 k^{\tau}\theta \mbox{d}\theta = \frac{1}{2}\delta_{1,k^{\tau}} 
\label{eq:cyy2}
\\
c_{XY}(\tau) &=& \frac{1}{2\pi}\int_{0}^{2\pi}\cos\theta \sin
 k^{\tau}\theta \mbox{d}\theta = 0 
\label{eq:cxy2}
\\
c_{YX}(\tau) &=& \frac{1}{2\pi}\int_{0}^{2\pi}\sin\theta \cos
 k^{\tau}\theta \mbox{d}\theta = 0 
\label{eq:cyx2}
\end{eqnarray}
These are extensions of Chebyshev maps derived by Geisel and Fairen to 
the two-dimensional map~\cite{Geisel}. Therefore, the auto-correlations
of $x$ and $y$ decay 0 for $\tau \geq 1$, and the cross-correlations
between $x$ and $y$ are zero. Furthermore, the correlation between $z_t$
and $\overline{z_{t+\tau}}$, where $\overline{\cdot}$ is denoted as the
complex conjugate of $\cdot$, is also zero,
\begin{equation}
\lim_{T\rightarrow\infty}\frac{1}{T}\sum_{t=0}^{T-1}z_t z_{t+\tau} =
 c_{XX}(\tau) - c_{YY}(\tau) +
 \sqrt{-1}\bigl(c_{XY}(\tau)+c_{YX}(\tau)\bigr) = 0.
\end{equation}

Note that Eqs. (\ref{eq:cxx2}) to (\ref{eq:cyx2}) are derived by means
of the permutability of $z^k$ and the orthogonality between $P_k(x,y)$
and $Q_k(x,y)$. Clearly, from Eq. (\ref{eq:2dmap}) we can prove the
permutability of $z^k$ such as $(z^k)^l = z^{kl}$. For $k \geq 1$ and $l
\geq 1$, we also have the orthogonal relations among $P_k(x,y)$ and
$Q_k(x,y)$
\begin{eqnarray}
\nonumber
\int_{-1}^{1}\mbox{d}x \int_{-1}^{1}\mbox{d}y P_k(x,y)P_l(x,y)\rho_{XY}(x,y) &=&
 \frac{1}{2\pi} \int_{0}^{2\pi}\cos(k\theta)\cos(l\theta)\mbox{d}\theta =
 \frac{1}{2}\delta_{k,l},
\\
\\
\nonumber
\int_{-1}^{1}\mbox{d}x \int_{-1}^{1}\mbox{d}y Q_k(x,y)Q_l(x,y)\rho_{XY}(x,y) &=&
 \frac{1}{2\pi} \int_{0}^{2\pi}\sin(k\theta)\sin(l\theta)\mbox{d}\theta =
 \frac{1}{2}\delta_{k,l},
\\
\\
\nonumber
\int_{-1}^{1}\mbox{d}x \int_{-1}^{1}\mbox{d}y Q_k(x,y)P_l(x,y)\rho_{XY}(x,y) &=&
 \frac{1}{2\pi} \int_{0}^{2\pi}\sin(k\theta)\cos(l\theta)\mbox{d}\theta =
0
\\
\end{eqnarray}

\section{Simultaneous higher order covariance}
\label{sec:sim-high}

Let us consider the characteristic function of the simultaneous joint density
$\rho_{XY}(x,y)$, defined as
\begin{eqnarray}
\nonumber
\Phi(u,v) &=& \lim_{T\rightarrow\infty}
\frac{1}{T}\sum_{t=0}^{T-1}e^{\sqrt{-1}(ux_t+vy_t)} \\
   &=& \int_{-\infty}^{\infty}
    \int_{-\infty}^{\infty}e^{\sqrt{-1}(ux+vy)}\rho_{XY}(x,y)\mbox{d}x\mbox{d}y.
\label{eq:characteristic}
\end{eqnarray}
Inserting Eq. (\ref{eq:rhoxy}) into Eq. (\ref{eq:characteristic}), we
have
\begin{eqnarray}
\nonumber
\Phi(u,v) &=&
 \frac{1}{2\pi}\int_{-\infty}^{\infty}e^{\sqrt{-1}(ux+vy)}
\frac{\delta(\sqrt{x^2+y^2}-1)}{\sqrt{x^2+y^2}}\mbox{d}x \mbox{d}y \\
\nonumber
&=& \int_{0}^{2\pi}\mbox{d}\theta \int_{0}^{\infty}r \mbox{d}r
e^{\sqrt{-1}(u\cos\theta + v\sin\theta)}\frac{\delta(r-1)}{2\pi r} \\
&=& \frac{1}{2\pi}\int_{0}^{2\pi}e^{\sqrt{-1}(u\cos\theta +
 v\sin\theta)}\mbox{d}\theta = J_{0}^{1,1}(u,v),
\label{eq:phi-uv}
\end{eqnarray}
where $J_{n}^{p,q}(u,v)$ is defined as
\begin{equation}
J_n^{p,q}(u,v) =
 \frac{1}{2\pi}\int_{0}^{2\pi}e^{\sqrt{-1}(u\cos(p\theta)+
 v\sin(q\theta)-n\theta)} \mbox{d}\theta.
\end{equation}
This is similar to the two-dimensional Bessel function which was studied by
Korsch et al.~\cite{Korsch}, however, it is a bit different from it. They
define the two-dimensional Bessel functions with three integer indices $n$,
$p$, and $q$ as 
\begin{equation}
\hat{J}_{n}^{p,q}(u,v) = \frac{1}{2\pi}\int_{-\pi}^{\pi}
 e^{\sqrt{-1}(u\sin(p\theta)+u\sin(q\theta)-n\theta)} \mbox{d}\theta
\label{eq:Bessel2}
\end{equation}
In his definition, the two-dimensional Bessel function consists of two sine
functions. However, in our definition this consists of cosine and sine
functions.

Clearly, both the two-dimensional Bessel functions satisfy
\begin{eqnarray}
J_0^{1,1}(u,0) = J_0(u), &\quad& J_0^{1,1}(0,v) = J_0(v), \\
\hat{J}_0^{1,1}(u,0) = J_0(u), &\quad& \hat{J}_0^{1,1}(0,v) = J_0(v),
\end{eqnarray}
where $J_n(u)$ is the Bessel function defined as
\begin{equation}
J_n(u) = \frac{1}{2\pi}\int_{-\pi}^{\pi}e^{\sqrt{-1}(n\theta - u
 \sin\theta)} \mbox{d}\theta.
\end{equation}

In the one-dimensional case, Eq. (\ref{eq:phi-uv}) is equivalent to
the characteristic function of Chebyshev polynomials, which is derived
by Geisel and Fairen~\cite{Geisel}. We can further expand $\Phi(u,v)$ in
terms of $u$ and $v$,
\begin{eqnarray}
\nonumber
\Phi(u,v) &=& \frac{1}{2\pi}\sum_{n=0}^{\infty}\frac{(\sqrt{-1})^n}{n!}\int_0^{2\pi}(u\cos\theta
 + v\sin\theta)^n \mbox{d}\theta \\
\nonumber
&=& \frac{1}{2\pi}\sum_{n=0}^{\infty}\frac{(\sqrt{-1})^n}{n!}\sum_{m=0}^n
\left(\begin{array}{c}n \\ m\end{array}\right)u^m v^{n-m}
\int_{0}^{2\pi}\cos^m \theta \sin^{n-m}\theta \mbox{d}\theta.
\end{eqnarray}
Therefore, we have
\begin{eqnarray}
\nonumber
\langle X^m Y^{n-m} \rangle &=&
 \lim_{T\rightarrow\infty}
 \frac{1}{T}\sum_{t=0}^{T-1}x^m_{t}y^{n-m}_{t}\\
\nonumber
&=& \int_{-\infty}^{\infty} \mbox{d}x
 \int_{-\infty}^{\infty} \mbox{d}y x^m y^{n-m}\rho_{XY}(x,y) \\
&=& \frac{1}{2\pi}\int_{0}^{2\pi}\cos^m \theta \sin^{n-m}\theta
 \mbox{d}\theta. \quad (0 \leq m \leq n).
\end{eqnarray}

We also have the equality
\begin{equation}
\int_{0}^{\pi/2} \cos^{2p-1}\theta \sin^{2q-1}\theta \mbox{d}\theta =
 \frac{1}{2}B(p,q) = \frac{1}{2}\frac{\Gamma(p)\Gamma(q)}{\Gamma(p+q)},
\label{eq:beta2}
\end{equation}
where $B(a,b)$ denotes the beta function, defined as
\begin{equation}
B(a,b) = \int_{0}^{1}\tau^{a-1}(1-\tau)^{b-1}\mbox{d}\tau,
\end{equation}
and $\Gamma(a)$ represents the gamma function, defined as
\begin{equation}
\Gamma(a) = \int_0^{\infty}e^{-\tau}\tau^{a-1}\mbox{d}\tau.
\end{equation}
Inserting Eq. (\ref{eq:beta2}) into $p=m/2+1/2$ and $q=(n-m)/2+1/2$ and
using symmetry of cosine and sine functions and $\Gamma(n+1)=n!$, we obtain
\begin{equation}
\langle X^m Y^{n-m} \rangle = 
\left\{
\begin{array}{ll}
\frac{2\Gamma(\frac{n-m+1}{2})\Gamma(\frac{m+1}{2})}{2\pi\Gamma(\frac{n}{2}+1)}
 = \frac{(m-1)!!(n-m-1)!!}{n!!} & (n,m:\mbox{even}) \\
0 & (\mbox{otherwise})
\end{array}
\right..
\end{equation}

Hence, the characteristic function of $\rho_{XY}(x,y)$ is described as
\begin{equation}
\Phi(u,v) = \sum_{n=0}^{\infty}(-1)^n\sum_{m=0}^{n}\frac{(u^2)^{m}(v^2)^{n-m}}{(2m)!!(2n-2m)!!(2n)!!}.
\end{equation}
This is a natural extension of the Bessel function of degree 0
to the two-dimensional case,
\begin{equation}
J_0(z) = \sum_{r=0}^{\infty}\frac{(-z^2)^r}{(2r)!!(2r)!!}.
\end{equation}

Since we can further calculate the $m$-th order moment of $x_t$ and the
$n-m$-th order moment of $y_t$ as
\begin{eqnarray}
\nonumber
\langle X^m \rangle &=& \int_{-1}^{1}\mbox{d}x \int_{-1}^{1}\mbox{d}y x^m
 \rho_{XY}(x,y) \\
&=& \frac{1}{2\pi}\int_{0}^{2\pi}\cos^m\theta \mbox{d}\theta = 
\left\{
\begin{array}{ll}
\frac{(m-1)!!}{m!!} & (m:\mbox{even}) \\
0 & (m:\mbox{odd}) 
\end{array}
\right.,
\label{eq:xm}
\end{eqnarray}
and
\begin{eqnarray}
\nonumber
\langle Y^{n-m} \rangle &=& \int_{-1}^{1}\mbox{d}x \int_{-1}^{1}\mbox{d}y y^{n-m}
 \rho_{XY}(x,y) \\
&=& \frac{1}{2\pi}\int_{0}^{2\pi}\sin^{n-m}\theta \mbox{d}\theta = 
\left\{
\begin{array}{ll}
\frac{(n-m-1)!!}{(n-m)!!} & (n-m:\mbox{even}) \\
0 & (n-m:\mbox{odd}) 
\end{array}
\right.,
\label{eq:ym}
\end{eqnarray}
where $m!! = 2\cdot 4 \cdot 6 \cdots m$ for even $m$ and $m!! = 1\cdot
3\cdot 5\cdots m$ for odd $m$, we get
\begin{eqnarray}
\nonumber
\mbox{Cov}[X^m,Y^{n-m}] &=& \langle X^m Y^{n-m} \rangle - \langle X^m
 \rangle \langle Y^{n-m} \rangle \\
&=&
\left\{
\begin{array}{ll}
\frac{(m-1)!!(n-m-1)!!}{n!!}\Bigl[1-\frac{n!!}{m!!(n-m)!!}\Bigr] &
 (m,n:\mbox{even}) \\
0 & (\mbox{otherwise})
\end{array}
\right.
\label{eq:cov}
\end{eqnarray}
Here, we consider the negativity of even ordered moments. Hammersley
suggested that antithetic variables are effective for variance reduction
in Monte Carlo integrations~\cite{Hammersley}. The antithetic-variates method
permits estimates through the use of negative correlated random
variables faster than independent random variables. Let us confirm the
sign of Eq. (\ref{eq:cov}). We get
\begin{eqnarray}
\nonumber
1 - \frac{n!!}{m!!(n-m)!!} &=& 1 -
 \frac{(\frac{n}{2})!}{(\frac{m}{2})!(\frac{n-m}{2})!} \\
&=& 1 - \left(\begin{array}{c}\frac{n}{2} \\
	      \frac{m}{2}\end{array}\right) \leq 0,
\label{eq:inequality}
\end{eqnarray}
since from the definition of combination, we have
\begin{equation}
\left(\begin{array}{c}
\frac{n}{2} \\
\frac{m}{2}
\end{array}\right) =
\frac{(\frac{n}{2})!}{(\frac{m}{2})!(\frac{n-m}{2})!} \geq 1.
\end{equation}
The equality is satisfied if and only if $m=0$ or $m=n$. Note that
Eq. (\ref{eq:cov}) is independent of a value of $k.$

\begin{figure}[!ht]
\centering
\includegraphics[scale=0.8]{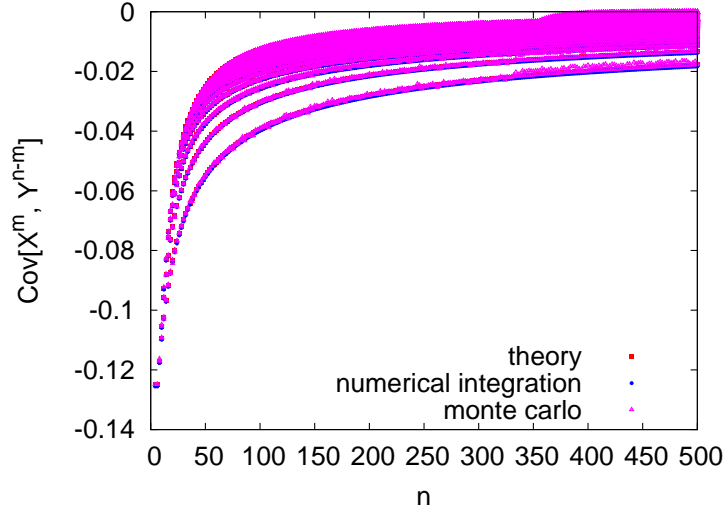}
\caption{The relationship between $n$ and $\mbox{Cov}[X^m,Y^{n-m}]$ for
 $k=2$.}
\label{fig:cov}
\end{figure}

Therefore, Eq. (\ref{eq:cov}) implies that $x_t$ and $y_t$ do not
have any correlations for the odd-ordered moments, however, do have a
negative covariance for the even-ordered moments. Fig. \ref{fig:cov}
shows the relationship between $n$ and $\mbox{Cov}[X^m,Y^{n-m}]$. It 
is confirmed that the covariance monotonically increases and approaches
to zero as $n$ increasing. 

Furthermore, we calculate covariance between $x_t^m$ and $x_t^{n-m}$,
and between $y_t^m$ and $y_t^{n-m}$. From Eqs. (\ref{eq:xm}) and
(\ref{eq:ym}), we have
\begin{eqnarray}
\nonumber
\mbox{Cov}[X^m,X^{n-m}]  &=& \mbox{Cov}[Y^m,Y^{n-m}] \\ 
\nonumber
&=&\left\{
\begin{array}{ll}
\frac{1}{2^n}\Bigl[
\left(\begin{array}{c}n \\ \frac{n}{2}\end{array}\right)
-
\left(\begin{array}{c}m \\ \frac{m}{2}\end{array}\right)
\left(\begin{array}{c}n-m \\ \frac{n-m}{2}\end{array}\right)

\Bigr] \geq 0 & (n,m:\mbox{even})\\
0 & (\mbox{otherwise})
\end{array}
\right.. \\
\label{eq:covxx-covyy}
\end{eqnarray}
The non-negativity of Eq. (\ref{eq:covxx-covyy}) is proven as follows.
Let us consider the case that $n$ is even. From
\begin{equation}
(1+x)^n = \Bigl\{(1+x)^{\frac{n}{2}}\Bigr\}^2,
\end{equation}
one has
\begin{equation}
\sum_{m=0}^n \left(\begin{array}{l}n \\ m \end{array}\right)x^m
= \Bigl(\sum_{m=0}^\frac{n}{2} \left(\begin{array}{l}\frac{n}{2} \\ m \end{array}\right)x^m\Bigr)^2
\end{equation}
Comparing $x^m$'s coefficient, we get the following inequality
\begin{equation}
\left(\begin{array}{l}n \\ m \end{array} \right) \geq \left(\begin{array}{l}\frac{n}{2} \\ \frac{m}{2} \end{array}\right)^2.
\end{equation}
Therefore, we obtain
\begin{equation}
\frac{1}{2^n}\Bigl[
\left(\begin{array}{c}n \\ \frac{n}{2}\end{array}\right)
-
\left(\begin{array}{c}m \\ \frac{m}{2}\end{array}\right)
\left(\begin{array}{c}n-m \\ \frac{n-m}{2}\end{array}\right)
\Bigr] = 
\frac{1}{2^n}\frac{\left(\begin{array}{l}n \\ \frac{n}{2}\end{array}\right)}{\left(\begin{array}{l}n \\ m \end{array}\right)}
\Bigl[\left(\begin{array}{l}n \\ m \end{array}\right) -
\left(\begin{array}{l}\frac{n}{2} \\
      \frac{m}{2}\end{array}\right)^2\Bigr] \geq 0.
\end{equation}

\section{Higher order covariance with lags}
\label{sec:lag-high}

More generally, we can introduce a characteristic function of the joint
density between
$x_{t+p}^m$ and $y_{t+q}^{n-m}$.
\begin{eqnarray}
\nonumber
\Psi_{XY}(u,v) &=&
 \lim_{T\rightarrow\infty}\frac{1}{T}\sum_{t=0}^{T-1}e^{\sqrt{-1}(ux_{t+p}+vy_{t+q})} \\
\nonumber
&=& \Bigl\langle \exp\bigl(\sqrt{-1}(u \underbrace{P_k\circ \cdots \circ
 P_k}_p(x,y) + v \underbrace{Q_k\circ \cdots \circ
 Q_k}_q(x,y))\bigr)\Bigr\rangle \\
\nonumber
&=& \int_{-1}^{1} \mbox{d}x \int_{-1}^{1}\mbox{d}y
\exp\bigl(\sqrt{-1}(u \underbrace{P_k\circ \cdots \circ
 P_k}_p(x,y) + v \underbrace{Q_k\circ \cdots \circ
 Q_k}_q(x,y))\bigr)\rho_{XY}(x,y) \\
&=& \frac{1}{2\pi}\int_{0}^{2\pi}e^{\sqrt{-1}(u\cos(k^p\theta)+
 v\sin(k^q\theta))} \mbox{d}\theta = J_0^{k^p,k^q}(u,v).
\end{eqnarray}
Similarly to $\Phi(u,v)$, from the expansion in terms of $u$ and $v$, we
obtain
\begin{eqnarray}
\nonumber
\lim_{T\rightarrow \infty}\frac{1}{T}\sum_{t=0}^{T-1}x_{t+p}^m
 y_{t+q}^{n-m} &=& \int_{-1}^{1}\mbox{d}x \int_{-1}^{1}\mbox{d}y
 \underbrace{P_k \circ \cdots \circ P_k}_p(x,y) \underbrace{Q_k \circ
 \cdots \circ 
 Q_k}_q(x,y)\rho_{XY}(x,y) \\
&=&
 \frac{1}{2\pi}\int_{0}^{2\pi}\cos^m(k^p\theta)\sin^{n-m}(k^q\theta)\mbox{d}\theta.
\end{eqnarray}
By using
\begin{eqnarray}
\nonumber
&& \cos^m(k^p\theta)\sin^{n-m}(k^q\theta) \\
\nonumber
&=& \frac{1}{2^m}(e^{\sqrt{-1}k^p\theta}+e^{-\sqrt{-1}k^p\theta})^m
\frac{1}{(2\sqrt{-1})^{n-m}}(e^{\sqrt{-1}k^q\theta}-e^{-\sqrt{-1}k^q\theta})^{n-m} \\
\nonumber
&=&
\frac{1}{2^n(\sqrt{-1})^{n-m}}\sum_{r=0}^m\sum_{s=0}^{n-m}(-1)^{n-m-s}\frac{m!}{r!(m-r)!}\frac{(n-m)!}{s!(n-m-s)!}e^{\sqrt{-1}[(2r-m)k^p+(2s-n+m)k^q]\theta}, \\
\end{eqnarray}
and 
\begin{equation}
\frac{1}{2\pi}\int_{0}^{2\pi}e^{\sqrt{-1}\alpha \theta}\mbox{d}\theta =
 \delta_{0,\alpha},
\end{equation}
we obtain
\begin{eqnarray}
\nonumber
&&\lim_{T\rightarrow
 \infty}\frac{1}{T}\sum_{t=0}^{T-1}x^{m}_{t+p}y_{t+q}^{n-m} \\
\nonumber
&=&
\left\{
\begin{array}{ll}
\frac{(-1)^{\frac{n-m}{2}}}{2^n}\sum_{r=0}^m\sum_{s=0}^{n-m}\frac{m!}{r!(m-r)!}\frac{(n-m)!}{s!(n-m-s)!}(-1)^{-s}\delta_{0,(2r-m)k^p+(2s-n+m)k^q}
 & \\
 & (m,n:\mbox{even}) \\
0 & (\mbox{otherwise})
\end{array}
\right.. \\
\end{eqnarray}
Since we further have
\begin{eqnarray}
\nonumber
\langle X_{t+p}^m \rangle &=& \int_{-1}^1\mbox{d}x \int_{-1}^1 \mbox{d}y
\Bigl[\underbrace{P_k \circ \cdots \circ P_k}_p(x,y)\Bigr]^m \rho_{XY}(x,y) \\
\nonumber
&=& \frac{1}{2\pi}\int_{0}^{2\pi}\cos^m(k^p \theta)\mbox{d}\theta \\
&=&
\left\{
\begin{array}{ll}
\frac{(m-1)!!}{m!!} & (m:\mbox{even}) \\
0 & (m:\mbox{odd}) 
\end{array}
\right.,
\end{eqnarray}
and
\begin{eqnarray}
\nonumber
\langle Y_{t+q}^{n-m} \rangle &=& \int_{-1}^1\mbox{d}x \int_{-1}^1 \mbox{d}y
\Bigl[\underbrace{Q_k \circ \cdots \circ Q_k}_q(x,y)\Bigr]^{n-m}\rho_{XY}(x,y) \\
\nonumber
&=& \frac{1}{2\pi}\int_{0}^{2\pi}\cos^{n-m}(k^q \theta)\mbox{d}\theta \\
&=&
\left\{
\begin{array}{ll}
\frac{(n-m-1)!!}{(n-m)!!} & (n-m:\mbox{even}) \\
0 & (n-m:\mbox{odd}) 
\end{array}
\right.,
\end{eqnarray}
we get
\begin{eqnarray}
\nonumber
&&\mbox{Cov}[X_{t+p}^m,Y_{t+q}^{n-m}] = \langle X_{t+p}^m Y_{t+q}^{n-m}
 \rangle - \langle X_{t+p}^m \rangle \langle Y_{t+q}^{n-m} \rangle \\
\nonumber
&=&
\left\{
\begin{array}{ll}
\frac{(-1)^{\frac{n-m}{2}}}{2^n}\sum_{r=0}^m\sum_{s=0}^{n-m}
\left(\begin{array}{c}m \\ r\end{array}\right)
\left(\begin{array}{c}n-m \\ s\end{array}\right)
(-1)^{-s}\delta_{0,(2r-m)k^p+(2s-n+m)k^q}\\
\qquad - \frac{(m-1)!!(n-m-1)!!}{m!!(n-m)!!}  & (m,n:\mbox{even}) \\
0 & (\mbox{otherwise})
\end{array}
\right..\\
\label{eq:high-cov}
\end{eqnarray}
Kohda et al. showed that the higher-order covariance of Chebyshev maps have
no correlation~\cite{Kohda:00}. We use their derivation in our
case. According to Kac's statistical independence~\cite{Kac}
when in Eq. (\ref{eq:high-cov}) 
\begin{equation}
(2r-m)k^p + (2s-n+m)k^q = 0, \quad (0 \leq r \leq m; 0 \leq s \leq n-m) 
\label{eq:integer-eq}
\end{equation}
holds for any $k^p$ and $k^q$ if and only if $r=m/2$ and $s=(n-m)/2$, $k^p$ and
$k^q$ are called linearly independent. Then $x^m_{t+p}$ and
$y^{n-m}_{t+q}$ are statistically independent~\cite{Kohda:00}.

Let consider the case that $m$ and $n$ are even. From elementary facts
about the theory of numbers, we know that 
\begin{equation}
N = k^{e} + r \quad (0 \leq r < k),
\end{equation}
where $N$ is a natural number, and $k$, $e$ and $r$ are non-negative integers.
In the case that $2r-m>0$, $2s-n+m<0$, and $p < q$ we have
\begin{eqnarray}
\nonumber
(2r-m)k^p + (2s-n+m)k^q &=& \{(2r-m) + (2s-n+m)k^{q-p}\}k^p \\
\nonumber
&=& \{(k^{e_1} + r')-(k^{e_2} + s')k^{q-p}\}k^p \\
&=& (k^{e_1}+r'-k^{e_2+q-p}-s'k^{q-p})k^p.
\end{eqnarray}
Therefore, if $[(2r-m)/k]=0$, $[(2s-n+m)/k]=0$, and $e_1= e_2+q-p$
hold then $(2r-m)k^p + (2s-n+m)k^q = 0$ is satisfied for integers
$r$ and $s$ other than $r=m/2$ and $s=(n-m)/2$. When $m>k$, and $n-m>k$, we
have $[(2r-m)/k]=0$ and $[(2s-n+m)/k]=0$. Therefore, $m \geq k^{e_1}$ 
and $n-m \geq k^{e_2}$ would be satisfied. Namely, when $n < k^{e_1} +
k^{e_2} = k^{e_2}(k^{q-p}+1)$, $x^m_{t+p}$ and $y^{n-m}_{t+q}$ are
statistically independent. This implies that $q-p$ goes infinity,
$x^{m}_{t+p}$ and $y^{n-m}_{t+q}$ become statistically independent in an
exponential manner.

\begin{figure}[!hbt]
\includegraphics[scale=0.5]{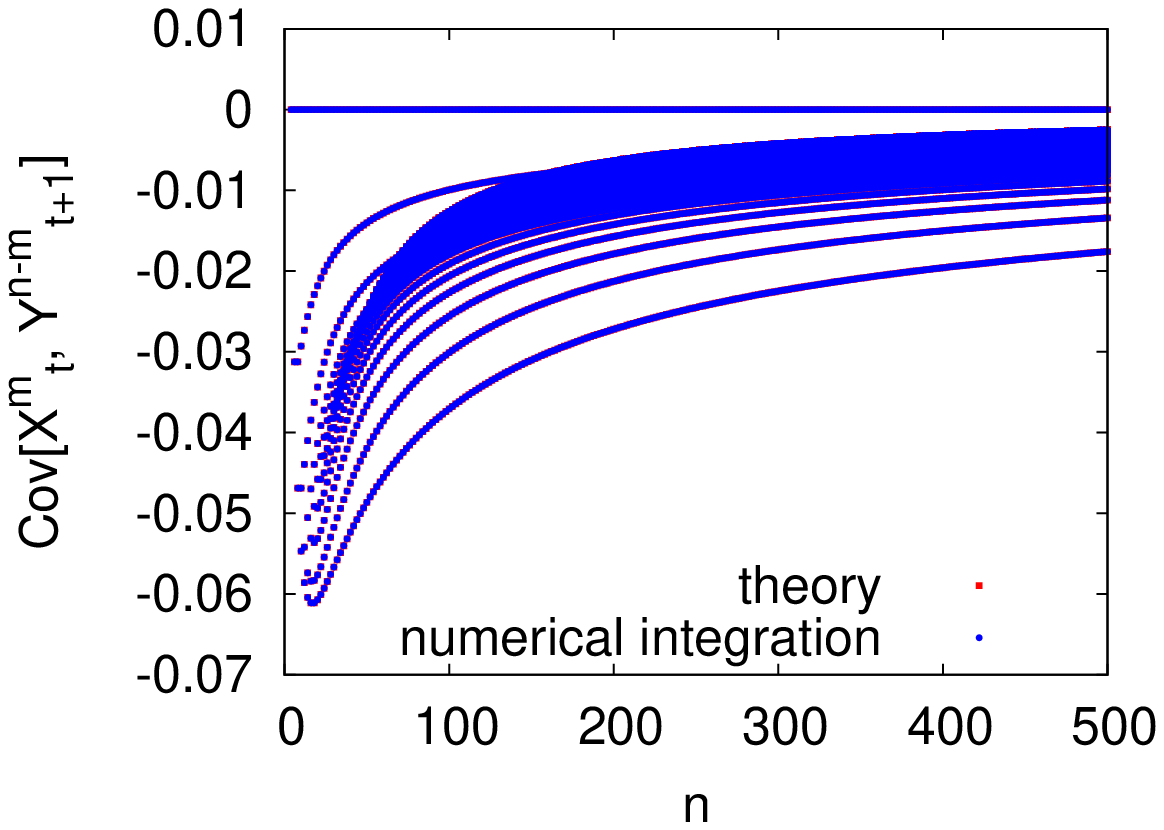}(a)
\includegraphics[scale=0.5]{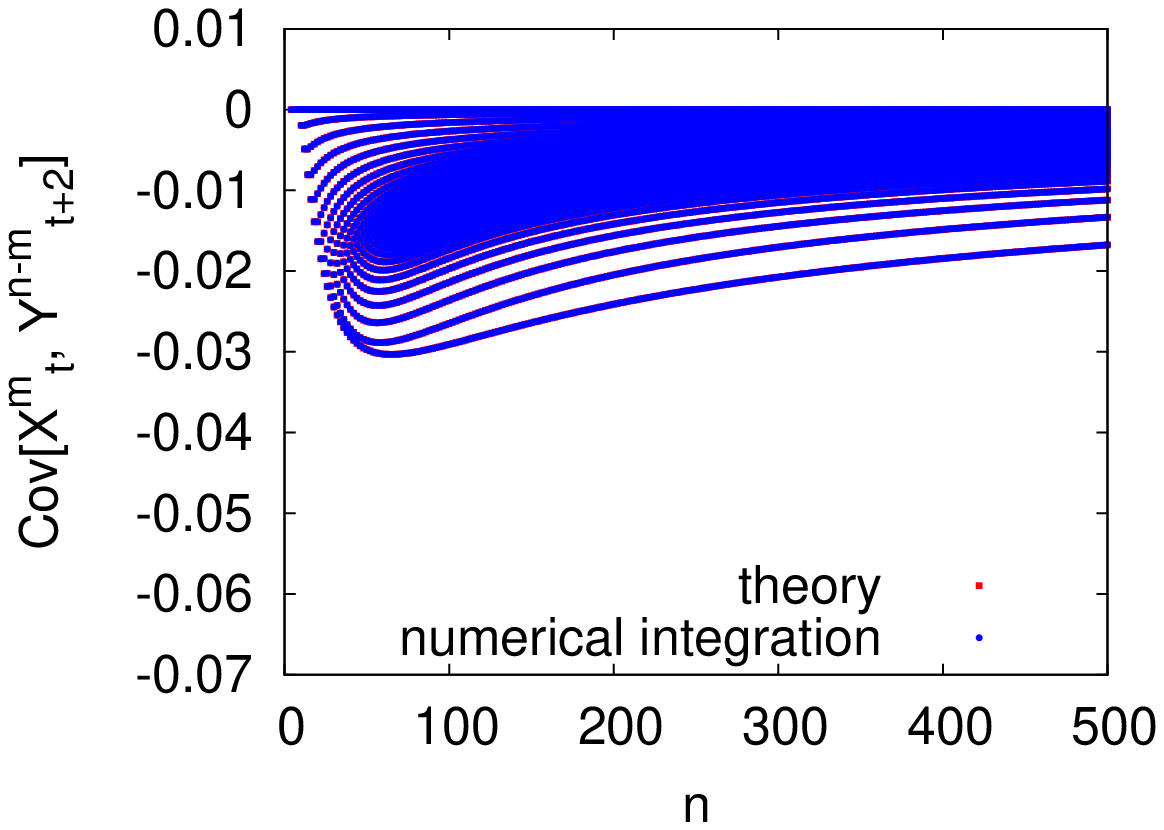}(b)
\includegraphics[scale=0.5]{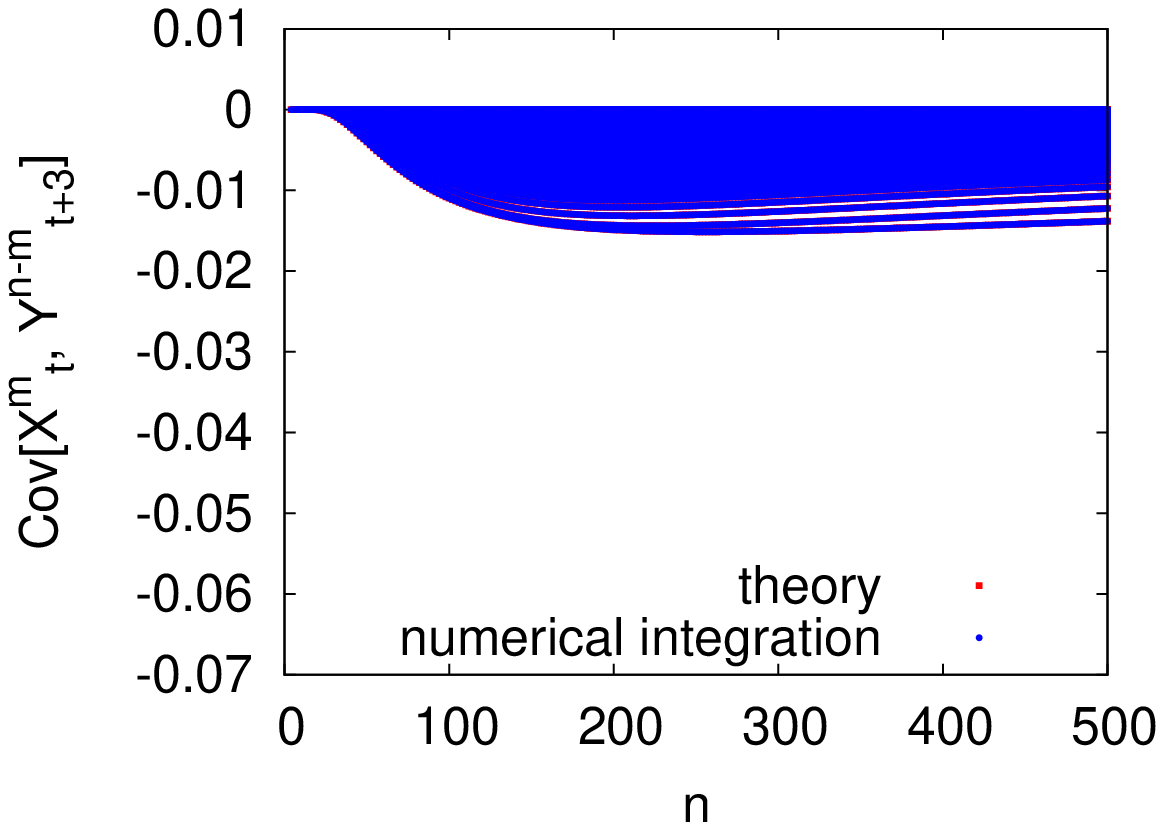}(c)
\includegraphics[scale=0.5]{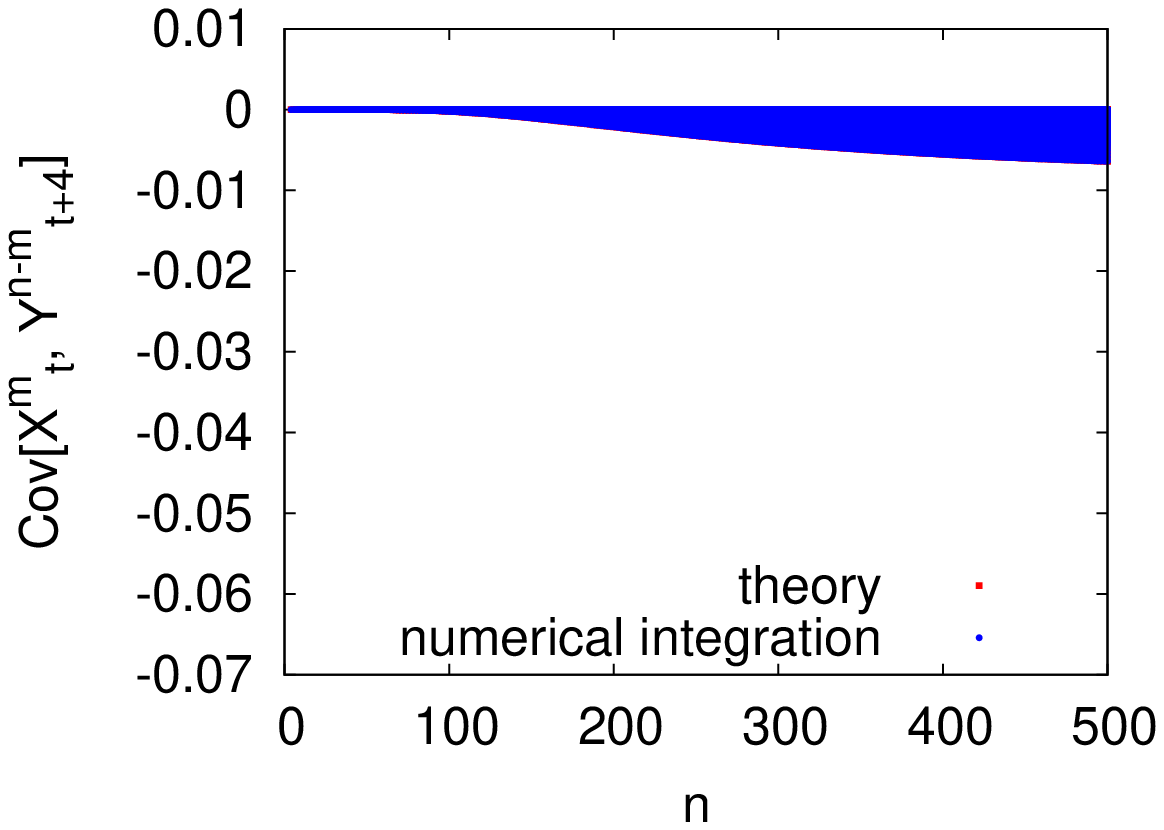}(d)
\includegraphics[scale=0.5]{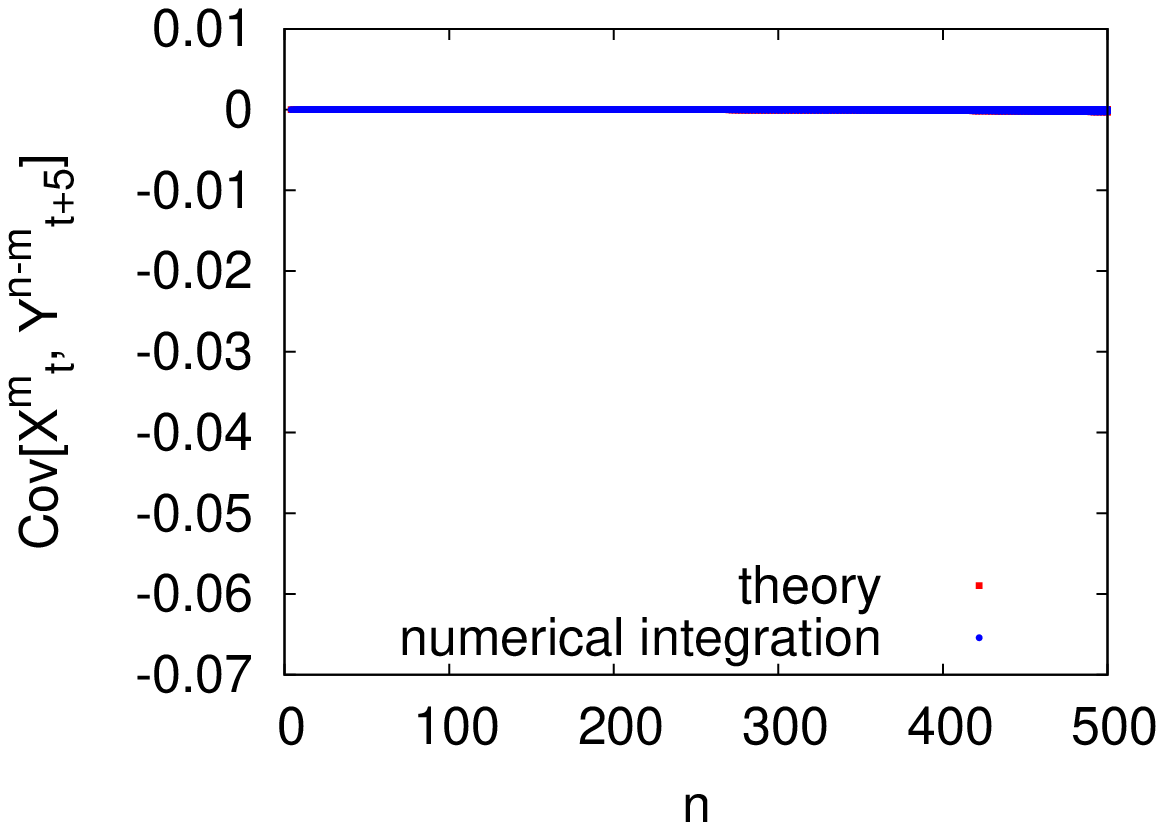}(e)
\includegraphics[scale=0.5]{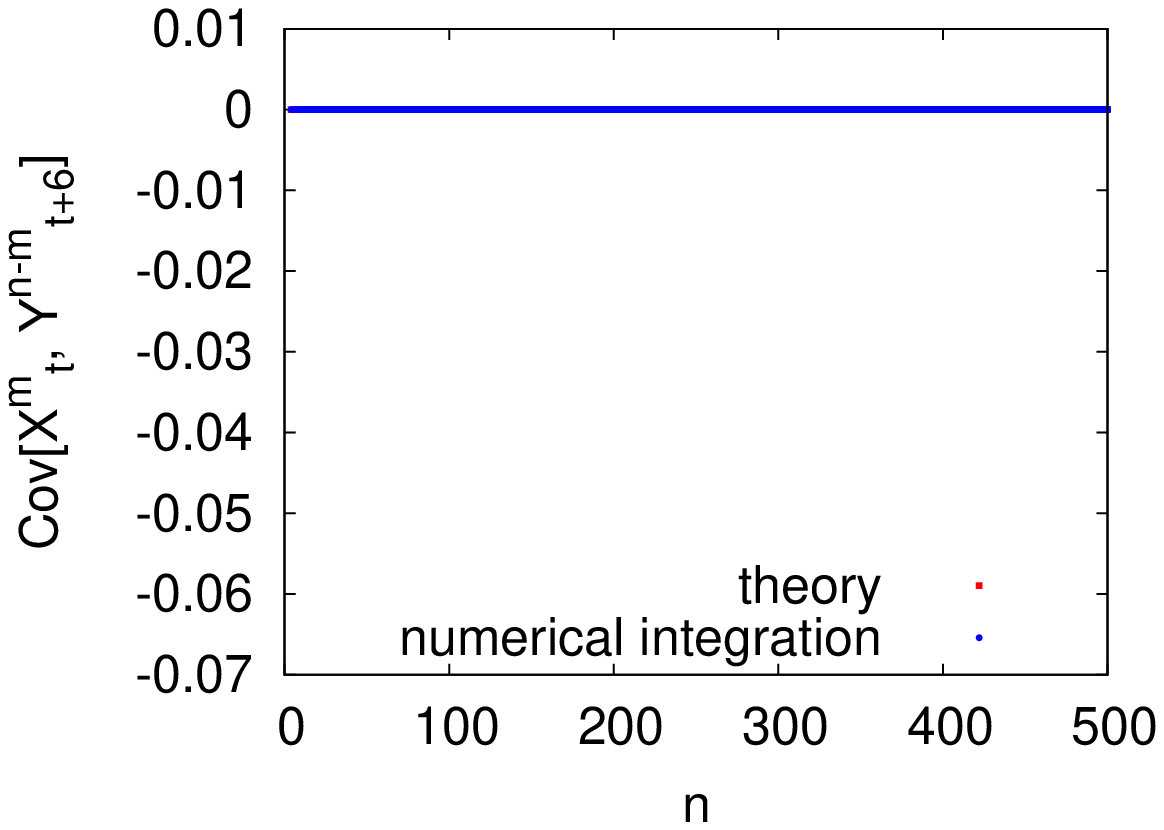}(f)
\caption{Scatter plots of $\mbox{Cov}[X^m_{t+p},Y^{n-m}_{t+q}]$ in terms
 of $n$ ($0 \leq m \leq n$) at $k=2$ and $p=0$, (a) $q=1$, (b) $q=2$, (c)
 $q=3$, (d) $q=4$, (e) $q=5$, and (f) $q=6$. Filled squares represent
 theoretical values, and filled circles values obtained from numerical
 integration.}
\label{fig:covpq}
\end{figure}

Fig. \ref{fig:covpq} shows $\mbox{Cov}[X^m_{t+p},Y^{n-m}_{t+q}]$ for
$(p,q)=(0,1)$, $(0,2)$, $(0,3)$, $(0,4)$, $(0,5)$, and $(0,6)$. As shown
in figures, we found that the covariances decrease as $|p-q|$
increasing. The range of the covariances approach to zero as $q$ increasing.

Obviously, Eq. (\ref{eq:integer-eq}) has solutions $r = m/2$ and $s =
(n-m)/2$. A sum of the contributions for $r=m/2$ and $s=(n-m)/2$ in
Eq. (\ref{eq:high-covyy}) is equivalent to
$\frac{(m-1)!!(n-m-1)!!}{m!!(n-m)!!}$. Since
$\mbox{Cov}[X_{t+p}^m,Y_{t+q}^{n-m}]$ is less than zero from the
numerical simulation, for solutions other than $r= m/2$ and $s =
(n-m)/2$ of Eq. (\ref{eq:integer-eq}), it should satisfy that a sum of 
negative contributions is greater than a sum of positive contributions.

We may consider two types of second-order characteristic functions with
lags. Note that Geisel and Fairen~\cite{Geisel} considered a similar
second-order characteristic function for the Chebyshev maps. Their
characteristic function corresponds to $\Psi_{XX}(u,u)$ in our definition.
\begin{eqnarray}
\nonumber
\Psi_{XX}(u,v) &=& \lim_{T\rightarrow \infty}\frac{1}{T}\sum_{t=0}^{T-1}
 e^{\sqrt{-1}(u x_{t+p} + v x_{t+q})} \\
&=&
 \frac{1}{2\pi}\int_{0}^{2\pi}e^{\sqrt{-1}(u\cos(k^p\theta)+v\cos(k^q\theta))}
 \mbox{d}\theta \\
\nonumber
\Psi_{YY}(u,v) &=& \lim_{T\rightarrow \infty}\frac{1}{T}\sum_{t=0}^{T-1}
 e^{\sqrt{-1}(u y_{t+p} + v y_{t+q})} \\
&=&
 \frac{1}{2\pi}\int_{0}^{2\pi}e^{\sqrt{-1}(u\sin(k^p\theta)+v\sin(k^q\theta))}
 \mbox{d}\theta 
\end{eqnarray}
Similarly to $\Psi_{XY}(u,v)$, from the expansion in terms of $u$ and $v$, we
obtain
\begin{eqnarray}
\Psi_{XX}(u,v) &=& \sum_{n=0}^{\infty}\frac{(-1)^{\frac{n}{2}}}{n!}\sum_{m=0}^{n}
\left(\begin{array}{l} n \\ m \end{array} \right)
\langle X^m_{t+p}X_{t+q}^{n-m} \rangle u^m v^{n-m}, \\
\Psi_{YY}(u,v) &=& \sum_{n=0}^{\infty}\frac{(-1)^{\frac{n}{2}}}{n!}\sum_{m=0}^{n}
\left(\begin{array}{l} n \\ m \end{array}\right)
\langle Y^m_{t+p}Y_{t+q}^{n-m} \rangle u^m v^{n-m},
\end{eqnarray}
where
\begin{eqnarray}
\nonumber
\langle X^m_{t+p}X^{n-m}_{t+q} \rangle = 
\lim_{T\rightarrow \infty}\frac{1}{T}\sum_{t=0}^{T-1}x_{t+p}^m
 x_{t+q}^{n-m} &=& 
\frac{1}{2\pi}
\int_{0}^{2\pi}\cos^m(k^p\theta)\cos^{n-m}(k^q\theta)\mbox{d} \theta, \\
\nonumber
\langle Y^m_{t+p}Y^{n-m}_{y+q} \rangle = 
\lim_{T\rightarrow \infty}\frac{1}{T}\sum_{t=0}^{T-1}y_{t+p}^m
 y_{t+q}^{n-m} &=&
\frac{1}{2\pi} \int_{0}^{2\pi}\sin^m(k^p\theta)\sin^{n-m}(k^q\theta)\mbox{d} \theta.
\end{eqnarray}

By using
\begin{eqnarray}
\nonumber
&& \cos^m(k^p\theta)\cos^{n-m}(k^q\theta) \\
\nonumber
&=& \frac{1}{2^m}(e^{\sqrt{-1}k^p\theta}+e^{-\sqrt{-1}k^p\theta})^m
\frac{1}{2^{n-m}}(e^{\sqrt{-1}k^q\theta}+e^{-\sqrt{-1}k^q\theta})^{n-m} \\
\nonumber
&=&
\frac{1}{2^n}\sum_{r=0}^m\sum_{s=0}^{n-m}\frac{m!}{r!(m-r)!}\frac{(n-m)!}{s!(n-m-s)!}e^{\sqrt{-1}[(2r-m)k^p+(2s-n+m)k^q]\theta}, \\
\nonumber
&& \sin^m(k^p\theta)\sin^{n-m}(k^q\theta) \\
\nonumber
&=& \frac{1}{2\sqrt{-1})^m}(e^{\sqrt{-1}k^p\theta}-e^{-\sqrt{-1}k^p\theta})^m
\frac{1}{(2\sqrt{-1})^{n-m}}(e^{\sqrt{-1}k^q\theta}-e^{-\sqrt{-1}k^q\theta})^{n-m} \\
\nonumber
&=&
\frac{(-1)^\frac{n}{2}}{2^n}\sum_{r=0}^m\sum_{s=0}^{n-m}\frac{m!}{r!(m-r)!}\frac{(n-m)!}{s!(n-m-s)!}(-1)^{-r-s}e^{\sqrt{-1}[(2r-m)k^p+(2s-n+m)k^q]\theta},
\end{eqnarray}
therefore, we have
\begin{eqnarray}
\nonumber
&\lim_{T\rightarrow \infty}\frac{1}{T}\sum_{t=0}^{T-1}x_{t+p}^m
 x_{t+q}^{n-m} = \\ 
&\left\{
\begin{array}{ll}
\frac{1}{2^n}\sum_{r=0}^m\sum_{s=0}^{n-m}\frac{m!}{r!(m-r)!}\frac{(n-m)!}{s!(n-m-s)!}\delta_{0,(2r-m)k^p+(2s-n+m)k^q}
 & (m,n:\mbox{even}) \\
0 & (\mbox{otherwise})
\end{array}
\right., \\
\nonumber
&\lim_{T\rightarrow \infty}\frac{1}{T}\sum_{t=0}^{T-1}y_{t+p}^m
y_{t+q}^{n-m} = \\
&\left\{
\begin{array}{ll}
\frac{(-1)^{\frac{n}{2}}}{2^n}\sum_{r=0}^m\sum_{s=0}^{n-m}\frac{m!}{r!(m-r)!}\frac{(n-m)!}{s!(n-m-s)!}(-1)^{-r-s}\delta_{0,(2r-m)k^p+(2s-n+m)k^q}
 & (m,n:\mbox{even}) \\
0 & (\mbox{otherwise})
\end{array}
\right.. 
\end{eqnarray}
We further have
\begin{eqnarray}
\nonumber
&&\mbox{Cov}[X_{t+p}^m,X_{t+q}^{n-m}] = \langle X_{t+p}^m X_{t+q}^{n-m}
 \rangle - \langle X_{t+p}^m \rangle \langle X_{t+q}^{n-m} \rangle \\
&=&
\left\{
\begin{array}{ll}
\frac{1}{2^n}\sum_{r=0}^m\sum_{s=0}^{n-m}
\left(\begin{array}{c}m \\ r\end{array}\right)
\left(\begin{array}{c}n-m \\ s\end{array}\right)
\delta_{0,(2r-m)k^p+(2s-n+m)k^q} \\
\qquad - \Bigl(\frac{(m-1)!!}{m!!}\Bigr)^2 & (m,n:\mbox{even}) \\
0 & (\mbox{otherwise})
\end{array}
\right..\\
\label{eq:high-covxx}
\end{eqnarray}
A sum of contributions for $r=m/2$ and $s=(n-m)/2$ in
Eq. (\ref{eq:high-covxx}) is equivalent to
$(\frac{(m-1)!!}{m!!})^2$. If Eq. (\ref{eq:integer-eq}) has 
other solutions than $r= m/2$ and $s = (n-m)/2$, then the covariance
positively increases. Therefore, we could prove
$\mbox{Cov}[X_{t+p}^m,X_{t+q}^{n-m}] \geq 0$.

We also have
\begin{eqnarray}
\nonumber
&&\mbox{Cov}[Y_{t+p}^m,Y_{t+q}^{n-m}] = \langle Y_{t+p}^m Y_{t+q}^{n-m}
 \rangle - \langle Y_{t+p}^m \rangle \langle Y_{t+q}^{n-m} \rangle \\
&=&
\left\{
\begin{array}{ll}
\frac{(-1)^{\frac{n}{2}}}{2^n}\sum_{r=0}^m\sum_{s=0}^{n-m}
\left(\begin{array}{c}m \\ r\end{array}\right)
\left(\begin{array}{c}n-m \\ s\end{array}\right)
(-1)^{-r-s}\delta_{0,(2r-m)k^p+(2s-n+m)k^q} \\
\qquad - \Bigl(\frac{(n-m-1)!!}{(n-m)!!}\Bigr)^2 & (m,n:\mbox{even}) \\
0 & (\mbox{otherwise})
\end{array}
\right..
\label{eq:high-covyy}
\end{eqnarray}

\begin{figure}[!hbt]
\includegraphics[scale=0.8]{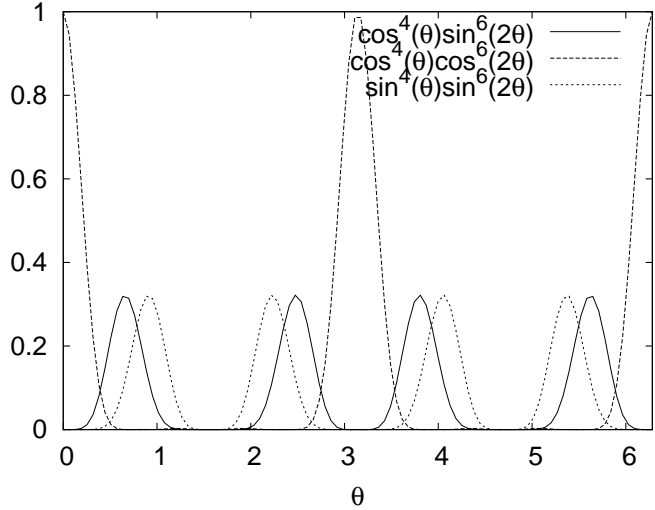}
\centering
\caption{The wave forms of 
$\cos^m(k^p\theta) \sin^{n-m}(k^q\theta)$, $\sin^m(k^p\theta)
\sin^{n-m}(k^q\theta)$, and $\cos^m(k^p\theta)
 \cos^{n-m}(k^q\theta)$ for $p=0$, $q=1$, $n=10$, and $m=4$.
}
\label{fig:sin-cos}
\end{figure}

\begin{figure}[!hbt]
\includegraphics[scale=0.5]{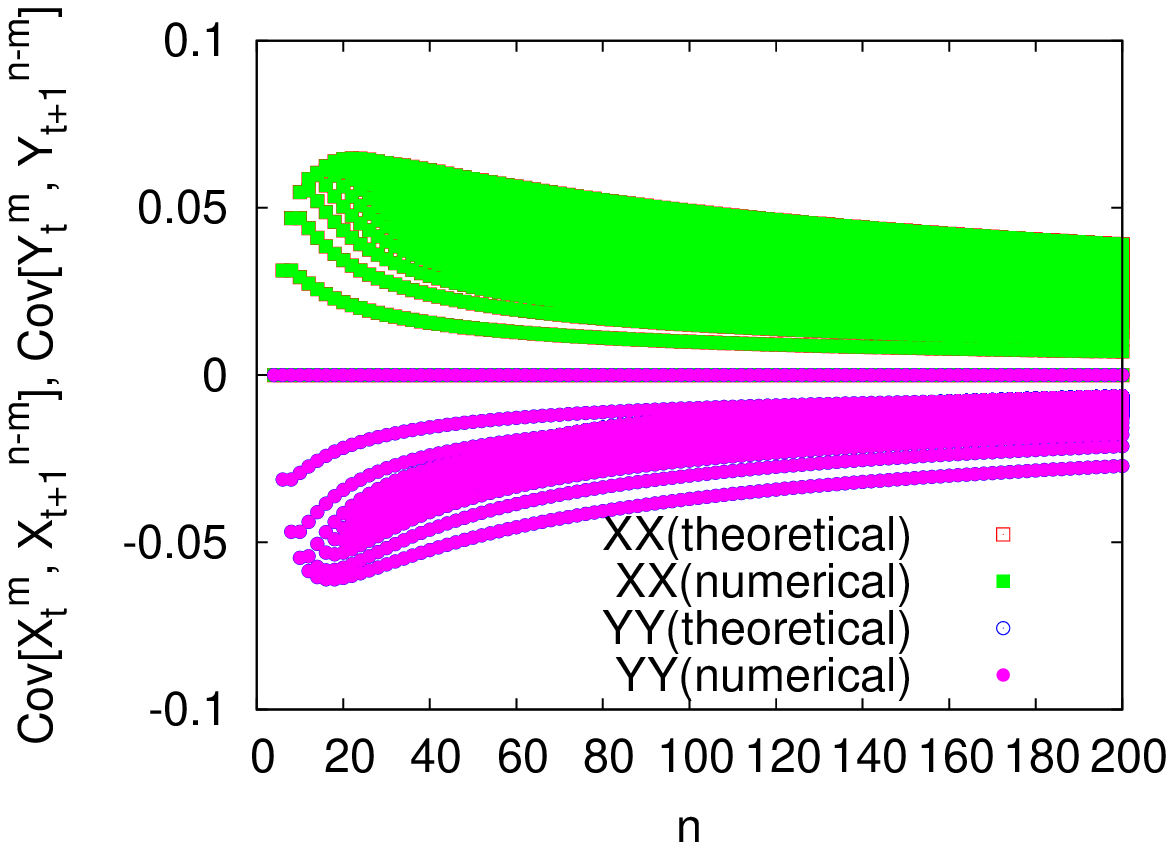}(a)
\includegraphics[scale=0.5]{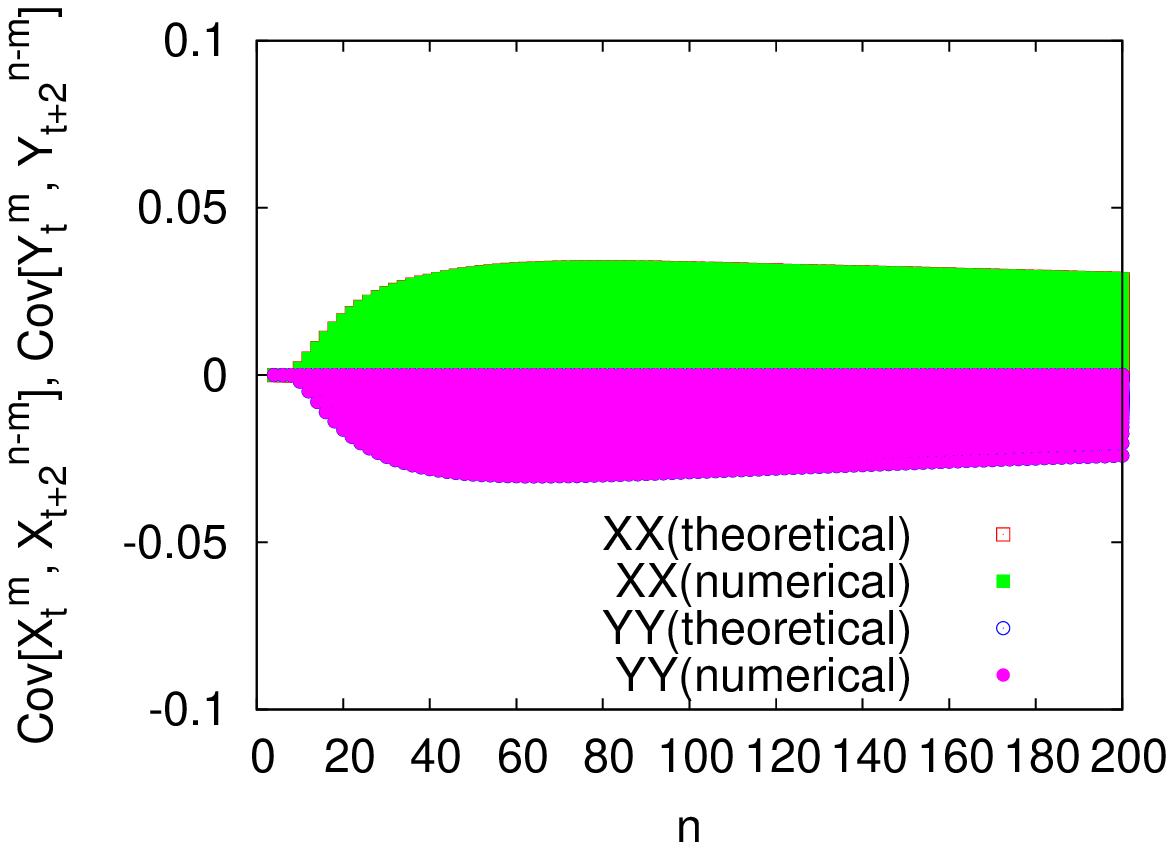}(b)
\includegraphics[scale=0.5]{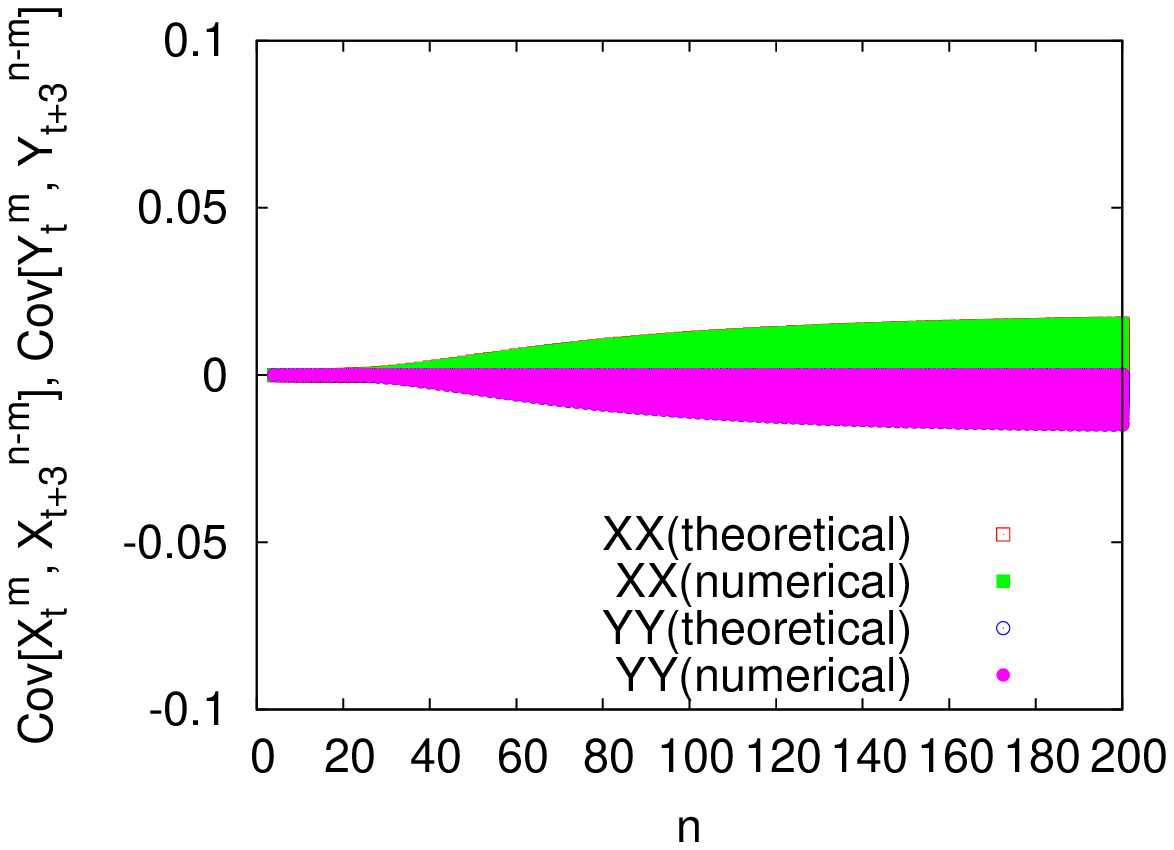}(c)
\includegraphics[scale=0.5]{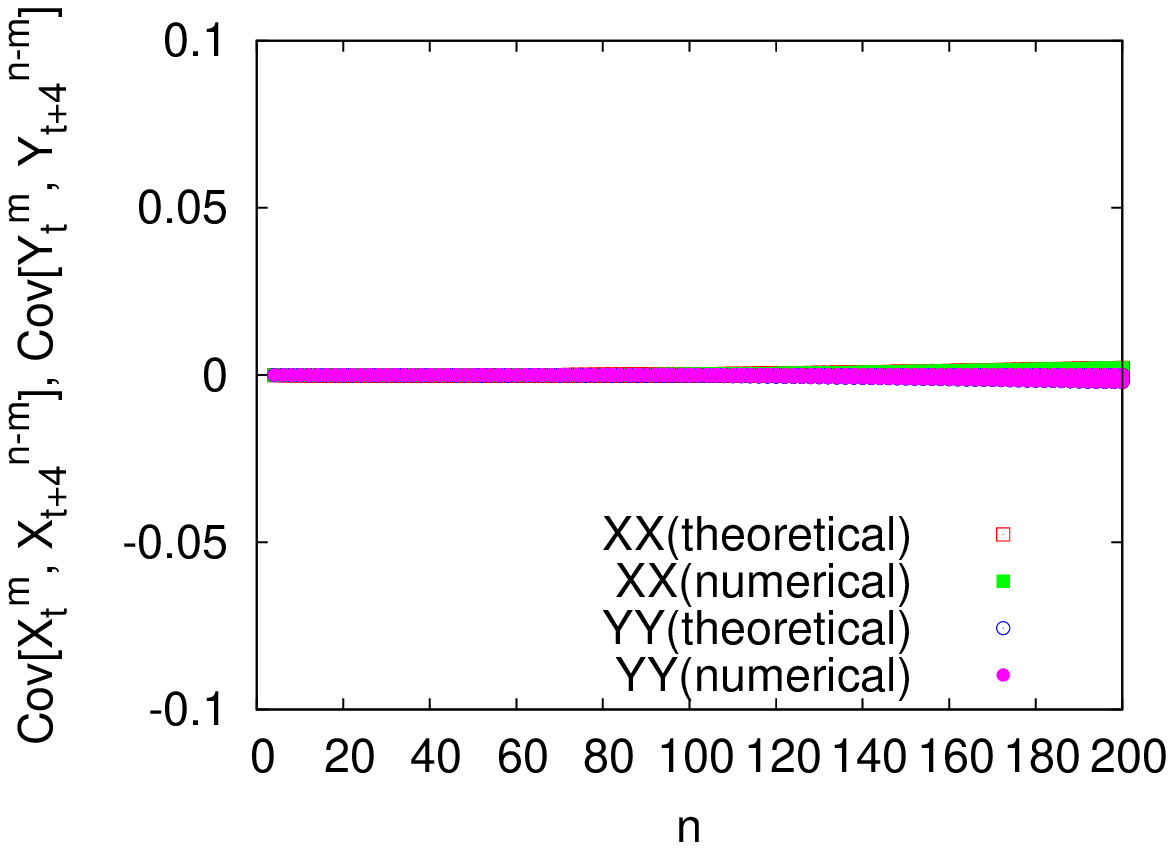}(d)
\includegraphics[scale=0.5]{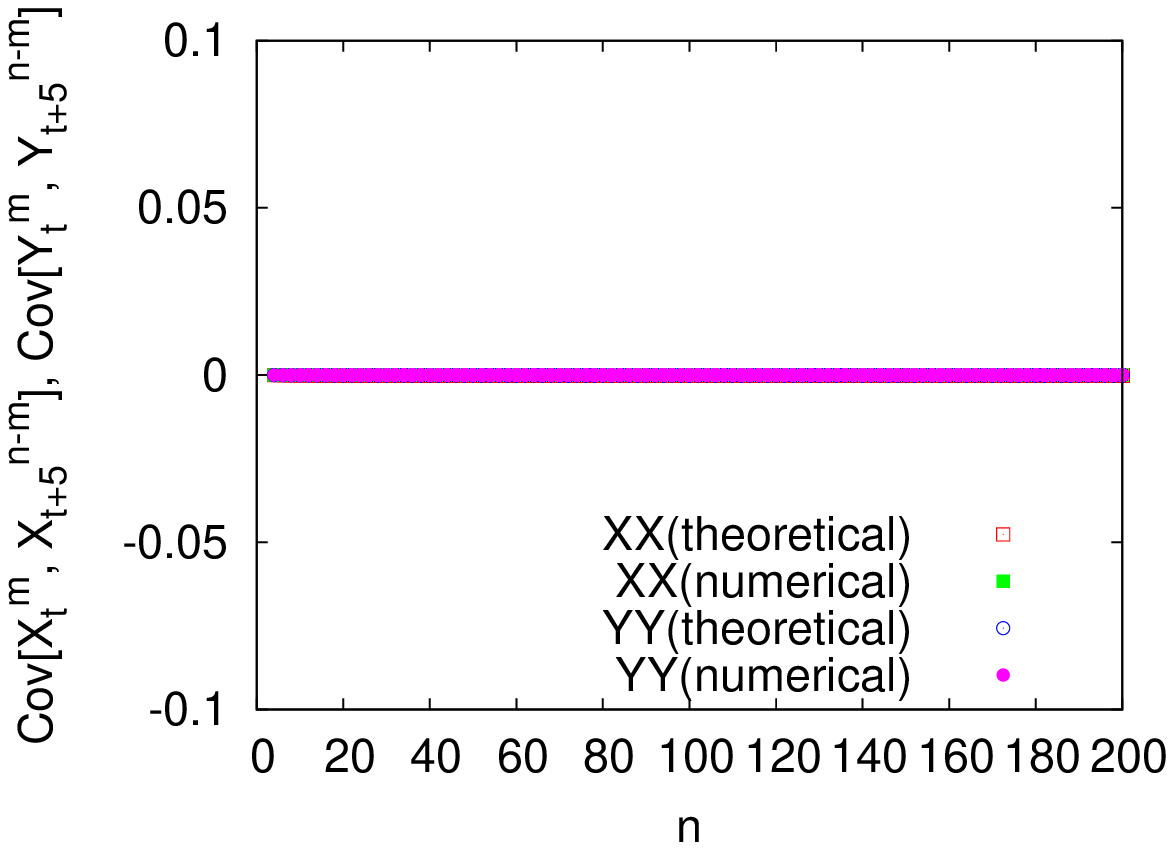}(e)
\includegraphics[scale=0.5]{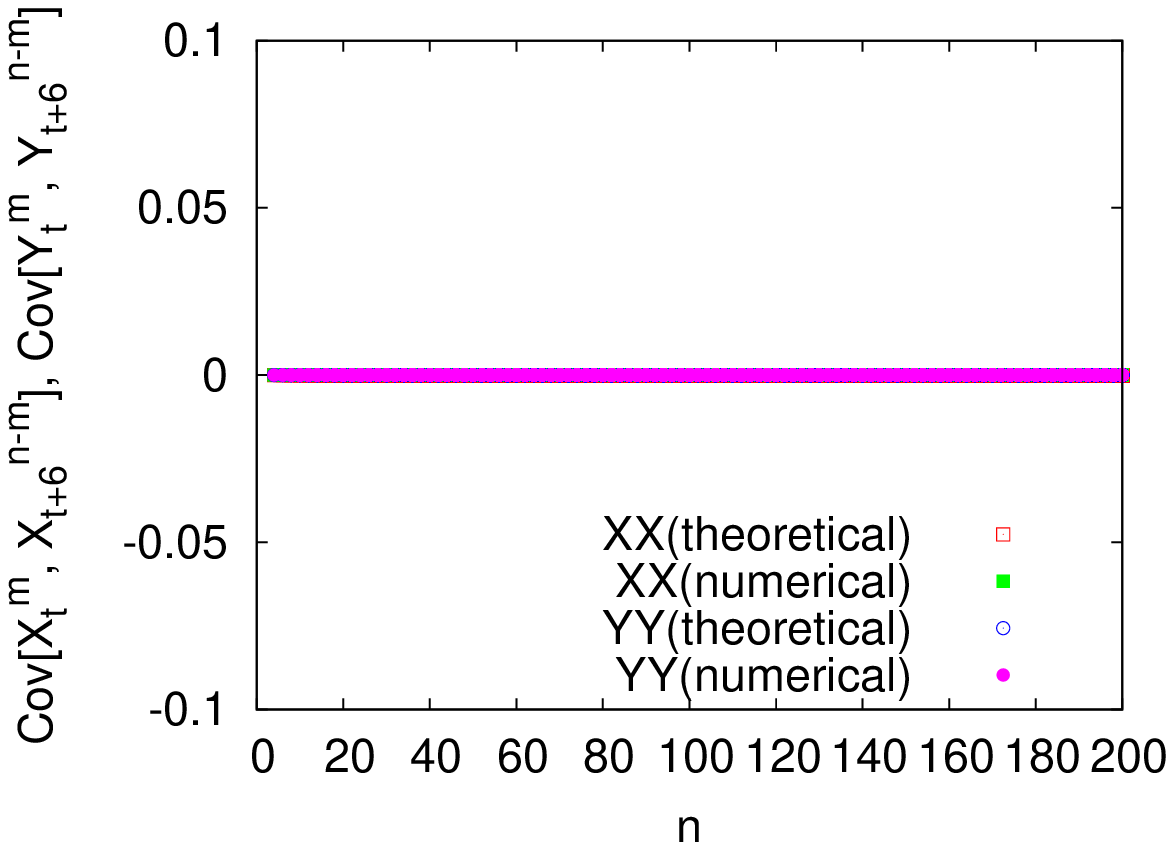}(f)
\caption{Scatter plots of $\mbox{Cov}[X^m_{t+p},X^{n-m}_{t+q}]$ and
 $\mbox{Cov}[Y^m_{t+p},Y^{n-m}_{t+q}]$ in terms of $n$ at $k=2$ and
 $p=0$, (a) $q=1$, (b) $q=2$, (c) $q=3$, (d) $q=4$, (e) $q=5$, and (f)
 $q=6$. Unfilled squares represent theoretical values of
 $\mbox{Cov}[X^m_{t+p},X^{n-m}_{t+p}]$, filled squares numerical values of
 $\mbox{Cov}[X^m_{t+p},X^{n-m}_{t+p}]$, unfilled circles theoretical
 values of $\mbox{Cov}[Y^m_{t+p},Y^{n-m}_{t+p}]$, and filled circles
 numerical values of $\mbox{Cov}[Y^m_{t+p},Y^{n-m}_{t+p}]$.}
\label{fig:covXXYYpq}
\end{figure}

Fig. \ref{fig:covXXYYpq} shows covariance between $X^m_{t+p}$ and
$X^{n-m}_{t+q}$, and between $Y^m_{t+p}$ and $Y^{n-m}_{t+q}$. It is
found that $\mbox{Cov}[X^m_{t+p},X^{n-m}_{t+q}]$ shows non-negative
values, and that $\mbox{Cov}[Y^m_{t+p},Y^{n-m}_{t+q}]$ shows
non-positive values. We found that 
$\mbox{Cov}[X^m_{t+p},Y^{n-m}_{t+q}]$ takes the same non-positive value as
$\mbox{Cov}[Y^m_{t+p},Y^{n-m}_{t+q}]$ for $p\neq q$
from Figs. \ref{fig:covpq} and \ref{fig:covXXYYpq}. The reason is
because $\cos^m(k^p\theta) \sin^{n-m}(k^q\theta)$ and $\sin^m(k^p\theta)
\sin^{n-m}(k^q\theta)$ have the same area to the x-axis, but 
$\cos^m(k^p\theta) \cos^{n-m}(k^q\theta)$ is different from them
as shown in Fig. \ref{fig:sin-cos}. 

A sum of the contributions for $r=m/2$ and $s=(n-m)/2$ in
Eq. (\ref{eq:high-covyy}) is equivalent to
$(\frac{(n-m-1)!!}{(n-m)!!})^2$. Since
$\mbox{Cov}[Y_{t+p}^m,Y_{t+q}^{n-m}]$ is less than zero from the 
numerical simulation, for solutions other than $r= m/2$ and $s =
(n-m)/2$ of Eq. (\ref{eq:integer-eq}), it should satisfy that a sum of
negative contributions is greater than a sum of positive contributions.

Therefore, it is suggested that $\Psi_{XX}(u,v) \neq \Psi_{YY}(u,v) \neq \Psi_{XY}(u,v)$
for $q\neq p$ from numerical simulation. This also implies that three
types of two-dimensional Bessel functions are not equivalent;
\begin{eqnarray}
J_{cc}^{p,q}(u,v) &=& \frac{1}{2\pi}\int_{0}^{2\pi}e^{\sqrt{-1}(u\cos(p\theta)+v\cos(q\theta))}\mbox{d}\theta, \\
J_{sc}^{p,q}(u,v) &=& \frac{1}{2\pi}\int_{0}^{2\pi}e^{\sqrt{-1}(u\sin(p\theta)+v\cos(q\theta))}\mbox{d}\theta, \\
J_{ss}^{p,q}(u,v) &=& \frac{1}{2\pi}\int_{0}^{2\pi}e^{\sqrt{-1}(u\sin(p\theta)+v\sin(q\theta))}\mbox{d}\theta. 
\end{eqnarray}

\section{Conclusion}
\label{sec:conclusion}
We studied two-dimensional chaotic maps on the unit circle, which is an
extension of the Chebyshev maps to two-dimensional map on the unit
circle. We examined correlational properties of this two-dimensional
chaotic map. We gave analytical forms of higher-order
moments. Furthermore, we derived the characteristic function of
both simultaneous and lagged ergodic densities. We found that these
characteristic functions are given by three types of two-dimensional
Bessel functions. We proved four theorems and proposed two conjectures
as follows:

\subsection*{Theorems:}
\begin{enumerate}
\item The higher-order covariances between $x_{t}$ and $y_{t}$ shows
      non-positive values for integers $n$ and $m \quad (0 \leq m \leq n)$:
      \begin{equation}
       \mbox{Cov}[X^m,Y^{n-m}] \leq 0.
      \end{equation}
\item The higher-order covariance between $x_{t}$ and $x_{t}$ shows 
      non-negative values for integer $n$ and $m \quad (0 \leq m \leq n)$:
      \begin{equation}
       \mbox{Cov}[X^m,X^{n-m}] \geq 0.
      \end{equation}
\item The higher-order covariance between $y_{t}$ and
      $y_{t}$ shows non-negative values for $n$ and $m \quad (0 \leq m
      \leq n)$: 
      \begin{equation}
       \mbox{Cov}[Y^m,Y^{n-m}] \geq 0.
      \end{equation}
\item The higher-order covariance between $x_{t+p}$ and $x_{t+q}$ $(p
      \neq q)$ shows non-negative values for integer $n$ and $m \quad (0
      \leq m \leq n)$:
      \begin{equation}
       \mbox{Cov}[X^m_{t+p},X^{n-m}_{t+q}] \geq 0.
      \end{equation}
\end{enumerate}

\subsection*{Conjectures:}
\begin{enumerate}
\item The higher-order covariances between $x_{t+p}$ and $y_{t+q}$ $(p
      \neq q)$ shows non-positive values for integers $n$ and $m \quad
      (0 \leq m \leq n)$:
      \begin{equation}
       \mbox{Cov}[X^m_{t+p},Y^{n-m}_{t+q}] \leq 0.
      \end{equation}
\item The higher-order covariance between $y_{t+p}$ and
      $y_{t+q}$ $(p \neq q)$ shows non-positive values for $n$ and $m
      \quad (0 \leq m 
      \leq n)$: 
      \begin{equation}
       \mbox{Cov}[Y^m_{t+p},Y^{n-m}_{t+q}] \leq 0.
      \end{equation}
\end{enumerate}

Therefore, we can generate antithetic sequences as $x_0, y_0, x_{1},
y_{1}, \ldots, x_{t}, y_{t}, \ldots$ or $y_{0}, y_{1}, y_{2}, \ldots,
y_t, \ldots$ obtained from Eq. (\ref{eq:map}). Asymmetric features
between cosine and sine functions were elucidated. Using the proposed
two-dimensional chaotic map, we can generate antithetic pseudo random 
sequences for Monte Carlo integration.

\end{document}